\documentclass[10pt,aps.prl,twocolumn,superscriptaddress,floatfix]{revtex4-2}

\usepackage{amsmath}
\usepackage{graphicx}
\usepackage{color}
\usepackage{caption}
\usepackage{subcaption}
\usepackage{float}
\usepackage{array}
\usepackage[margin=1.0in]{geometry}
\usepackage{ulem}
\usepackage{amssymb}
\usepackage{mathtools}

\begin{document}
\title{Neon tetra fish (Paracheirodon innesi) as farm-to-optical-table Bragg reflectors}
\author{D. Ryan Sheffield}
\affiliation{Florida Polytechnic University, Department of Physics, Lakeland, FL 33812, USA}
\author{Anthony Fiorito III}
\affiliation{Florida Polytechnic University, Department of Physics, Lakeland, FL 33812, USA}
\author{Hengzhou Liu}
\affiliation{Florida Polytechnic University, Department of Physics, Lakeland, FL 33812, USA}
\author{Michael Crescimanno}
\affiliation{Youngstown State University, Department of Physics, Youngstown, OH 44555, USA}
\author{Nathan J. Dawson}
\email{ndawson@floridapoly.edu}
\affiliation{Florida Polytechnic University, Department of Physics, Lakeland, FL 33812, USA}
\affiliation{Youngstown State University, Department of Physics, Youngstown, OH 44555, USA}
\affiliation{Department of Physics and Astronomy, Washington State University, Pullman, WA 99614, USA}
\date{\today}

\begin{abstract}
\noindent Iridophore networks in the skin of neon tetra fish are investigated for use as biologically sourced, tunable, Bragg reflector arrays. This paper reports on a method for immediate and fast post-processing of tissue to modify the structural color of iridophores found in the lateral color stripe. Conditions for fixation as well as the environment post-fixation to improve longevity of the structural color are also presented. Recent results from attempts to further increase the lifetime of post-mortem iridophore color through infiltration and embedding in low-acid glycol methacrylate are also discussed.
\end{abstract}

\maketitle

\vspace{1em}

\section{Introduction}
In 1978, REO Speedwagon released their seventh album titled, ``You Can Tune a Piano, but You Can't Tuna Fish,'' a clever play on words. Several researcher have since shown that you can, in fact, tune the iridescent color of a particular fish species.\cite{lythg82.01,cloth87.01,nagai89.01,lythg89.01,nagai92.01,yoshi11.01,gur15.01} The iridescence of fish skin is caused by specialized cells known as iridophores, which contain guanine crystal platelets surrounded by cytoplasm.\cite{land72.01} Silver-colored fish have iridophores with aperiodic photonic structures.\cite{dento71.01,zhao15.01,brady13.01,brady15.01} The large refractive index contrast between guanine crystals and cytoplasm results in broadband reflection of visible light.\cite{jorda12.01,dawso22.01} Many tropical fish species are quite colorful. Much of the color can be attributed to dye and pigment molecules in their chromatophores; however, some species with colorful iridescent markings have specialized iridophores with photonic crystals made from guanine crystal platelets periodically separated by layers of cytoplasm.\cite{kinos08.01,sun13.01,vigno16.01,gur17.01,gur20.01}

It is well-known that periodically layered, transparent materials reflect light over a range of wavelengths. A perfect bilayer reflector has quarter-wavelength optical thicknesses, $\lambda/4n_i$, with the reflection band at normal incidence centered at the free-space wavelength $\lambda$.\cite{perry65.01,orfan16.01} Here, $n_i$ are the low and high refractive indices. Many inorganic materials have been used in bilayer reflecting films due to 1) the wide assortment of materials that can result in large mismatches between the refractive indices of the two materials, $\Delta n = \left|n_2 - n_1\right|$, and 2) the development of high-precision methods to create uniform layered films such as sputtering and thermal evaporation. In the last few decades, many reflective films have been developed using organic optical materials and devices.\cite{kazmi07.01,scoto10.01,mao11.01,golde13.01} Some common polymers can be used to create reflective films which are cost effective.\cite{singe08.01,andre13.01,dawso23.01} Multilayer Bragg reflector films made from common polymers often require many layers due to the relatively low mismatch between their refractive index.\cite{singe08.01,lu20.01} Some specialty polymers and inorganic/organic hybrid materials with large refractive indices have been developed in recent times to be used in Bragg reflectors,\cite{manfr17.01,tavel20.01,athan21.01,jung21.01,zhang23.01} but most of these polymers are rather expensive which negates the argument for cost-effective solutions to reflective films. Additionally, fluoropolymers are the state-of-the-art polymers used as the low refractive index layer in polymer-based reflectors to achieve a relatively high mismatch in refractive indices.\cite{andre12.01,palo23.01}

\begin{figure*}[t]
\centering\includegraphics[scale=1]{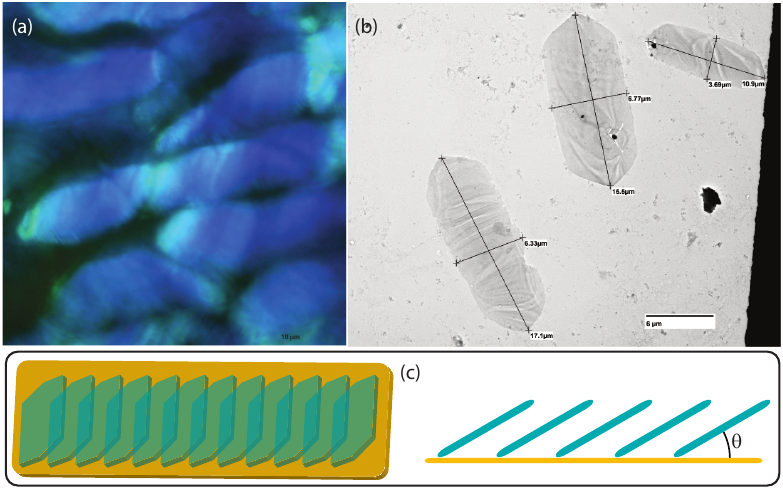}
\caption{(a) An optical microscope image of neon tetra iridophores. (b) TEM image of single guanine crystal platelets from neon tetra. (c) Cartoon illustrating the Venetian-blind model for the reflection band tunability of the neon tetra color stripe.}
\label{fig:fig1}
\end{figure*}

In this paper, we consider the use of biological photonic crystals as reflective materials. Consider the following question, how would you synthesize ethanol? The short answer is that you would not synthesize it because the most cost-effective method for producing ethanol is to simply let yeast do the work for you. In that same spirit, we asked ourselves a similar question regarding biological organic photonic crystals. In tropical reef systems there are brilliantly colored fish, some of which have quite beautiful iridescent colors. Here, we study the freshwater fish, neon tetra (Paracheirodon innesi) as a means to fabricate biological-based, microscopic, Bragg reflector arrays. The brightly colored iridophores form an array of reflectors as shown in Figure \ref{fig:fig1}, where previous evidence suggests the mechanism of color change is associated with the tilt angle of anchored guanine crystal platelets.\cite{gur15.01} This study focuses on three elements for the use of neon tetra iridophores as a means of fabricating low-cost Bragg reflector arrays: 1) the ease and effectiveness of tuning the reflection band, 2) the degradation of the Bragg reflectors over time, and 3) post-processing methods to extend the operational lifetime of the reflectors.

\section{Experimental methods}

This study involved materials sourced from live animals with protocol approved by the University of South Florida IACUC (IS00011971). Neon tetra fish were kept in a $40$ gallon breeder tank prior to experimentation. Each fish was euthanized by placing a specimen in a $0.3\,$wt.\% eugenol colloidal mixture with tank water for $60$ seconds. Each fish was washed for $10$ seconds in deionized (DI) water following euthanasia. Death was ensured by separating the spinal column at the base of the skull with a \#11 scalpel blade.

For tunability measurements, whole carcasses were bathed in a potassium monophosphate (KH$_{2}$PO$_{4}$) aqueous solution for $3$ minutes directly after euthanasia. The solutions were made by dissolving KH$_{2}$PO$_{4}$ salt at various concentrations in $18.2\,$M$\Omega$ DI water. The samples were descaled by lightly scraping the lateral sides with a razor blade, starting from the head and moving towards the tail. The reflectance spectrum was taken with a USB 4000 over a circular area with a diameter of $\sim 100\,\mu$m. White paper was used as the reflectance reference. Microscope images were taken with a BH-2 fluorescence microscope. A $24\,$V, $150\,$W halogen bulb and DC power supply was swapped out for the original mercury bulb and ballast. A $50/50$ beam splitter cube allowed for bright field images.

To prepare samples of single guanine platelets for imaging with a transmission electron microscope (TEM), carcasses were frozen and thawed followed by centrifuging at $10,000$ rpm for $10$ minutes. The supernatent was removed and $20\,\mu$L of DI water was added to the pellet in a conical centrifuge tube. The freezing, thawing, and centrifuging process was repeated two more times. The pellet was then placed in $20\,\mu$L of DI water for $1$ hour at room temperature. A droplet was placed on a UV-treated formvar grid for $1$ minute. The droplet was removed from the grid by touching filter paper to the edge of the grid/droplet. A DI water droplet was placed on the grid for $1$ minute and then removed again with filter paper. No negative staining was used for the TEM measurements. TEM images were taken by a Hitachi HT7700.

Biological samples fixed with mixtures of ethanol dehydrate. For the tunable iridophores found on the color stripe of neon tetra fish, fixing in ethanol removes the separation between adjacent guanine crystals. Thus, fixing neon tetra iridophores in ethanol results in a pale white color stripe -- a process which we have not yet found a way to reverse. Fixation with formalin is an excellent way to preserve some properties of biological samples and reduce degradation. A $10$\% stock solution of formalin contains $1$\% methanol and $3.7$\% formaldehyde. Formaldehyde, however, reacts with primary amine functional groups when fixing biological tissue.\cite{thava12.01} The photonic structures in fish iridophores are composed of guanine crystals, where the primary amine of guanine reacts with formaldehyde. This process is slow, where a $10$\% stock solution of formalin diluted with DI water sufficiently fixed the tissue to reduce naturally occurring degradation while maintaining the integrity of the photonic crystal.

For the fixation process used in degradation studies, carcasses were initially placed in a $5$\% formalin solution with $0.15\,$M KH$_{2}$PO$_{4}$. After $10$ minutes, the sample was placed in a formalin/KH$_{2}$PO$_{4}$ aqueous solution with the same $0.15\,$M KH$_{2}$PO$_{4}$ concentration, but the formalin concentration was reduced to $3$\%. After $1$ hour, the whole carcass was removed from solution and dorsally sectioned. The sample was placed in another $3$\% formalin solution with the same KH$_{2}$PO$_{4}$ concentration for $1.5$ hours. Most of the scales are removed during the fixation process, but a razor blade was used to gently remove any remaining scales after fixation.

Samples were also plastized to fabricated solid-state photonic structures along the lateral color stripe. The samples were fixed in the same manner as previously described. After the final fixation step, the halved carcasses were infiltrated with $85$\% low-acid GMA for $30$ minutes. The sample was then transferred to $97$\% low-acid GMA and infiltrated for $30$ minutes followed by a second $30$-minute infiltration of $97$\% low-acid GMA. Afterwards, it was placed in a gel capsule with undiluted low-acid GMA. A few drops of prepolymerized low-acid GMA was added to the capsule and sealed. The capsule was illuminated with ultraviolet light overnight, which resulted in embedded neon tetra tissue. Prepolymerization of low-acid GMA was achieved by dissolving benzoyl peroxide in a mixture of low-acid GMA, water, and butyl methacrylate, then filling a flask with a small layer of the solution. The flask was heated on a hot plate with constant swirling until the the solution began to yellow and thicken, then dunked into an ice bath with constant swirling to stop the solution from fully polymerizing.

\section{Results and discussion}

\begin{figure}[t]
\centering\includegraphics[scale=1]{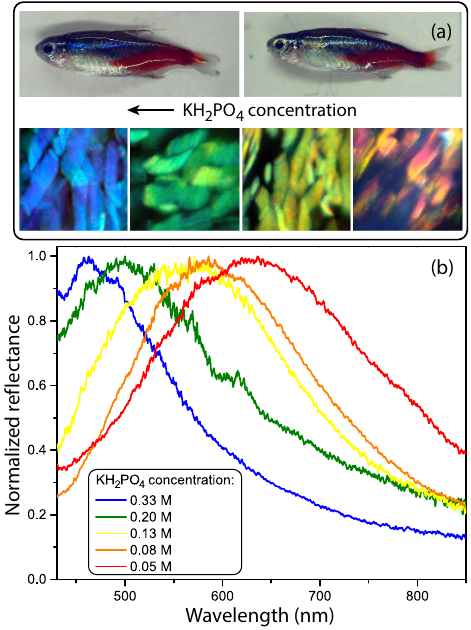}
\caption{(a) Pictures and microscope images of neon tetra fish soaked in high-concentration (left) and low-concentration (right) KH$_{2}$PO$_{4}$ solutions. (b) Normalized reflection spectrum after soaking in solutions with various KH$_{2}$PO$_{4}$ concentrations.}
\label{fig:fig2}
\end{figure}

Reflection measurements were taken from fish over a $\sim 1\,$mm diameter spot size. The fiber interface was $\sim 1\,$mm from the reference and sample surface. The white paper reference is ideal for diffuse reflections, whereas a polished aluminum sample is ideal for specular reflections. We tested two rough aluminum alloy samples to use as references. The sample that came from an extruded $1/16$" thick aluminum flat had visual color lines which exhibited location-specific deformations in the reference spectrum. The unpolished surface from cast aluminum stock showed shallow scallops in the reflection band when checked against white paper and extruded aluminum, which was assumed to be caused by interference effects from a thin film coating over the surface. The white paper was ideally suited as a diffuse reflection surface in the spectral range of $430\,$nm to $850\,$nm. Because the light source was chosen to be halogen, minimal UV light illuminated the area of interest which reduced any significant influence to the spectrum from optical brighteners. All reflection measurements were normalized.

The results from Figure \ref{fig:fig2} clearly show a KH$_{2}$PO$_{4}$ concentration-dependent tunability of the peak reflectance for arrays of neon tetra iridophores. The width of the reflection band increases as the peak reflectance moves towards the red. There are two reasons for this broadening, both of which can be observed in the microscope images in Figure \ref{fig:fig2}. When the reflectance is chemically tuned to the red end of the visible spectrum, single iridophores can reflect a broad range of colors, which indicates heterogeneous spacing between guanine platelets. This heterogeneous broadening over the surface of a single iridophore is also seen in microscope images for iridophores tuned to the blue end of the spectrum; however, the heterogeneity inside over a single iridophore is much less prominent. The second source of broadening is caused by the heterogeneous peak reflectance wavelength between neighboring iridophores. As observed in the microscope images in Figure \ref{fig:fig2}, neon tetra color stripes that have been chemically tuned to the red show a broader range of structural color in clusters of iridophores whereas blue-tuned color stripes exhibit a smaller degree of color heterogeneity among neighboring iridophores.

\begin{figure}[t]
\centering\includegraphics[scale=0.82]{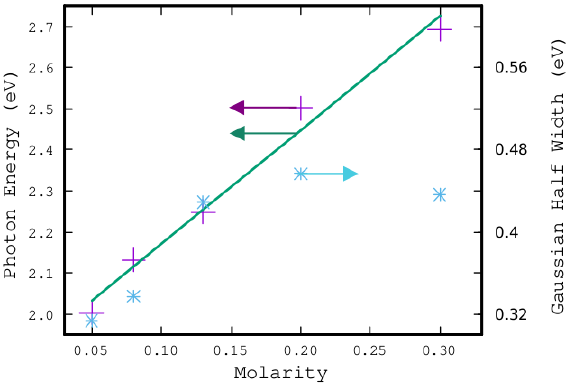}
\caption{A plot of the peak reflectivity photon energy versus the KH$_{2}$PO$_{4}$ molarity with trend line (purple crosses, green line), along with the width of the reflectivity spectrum at that concentration (cyan $\mathrlap{\times}+$) all for the data shown. The difference in these data trends is evidence against a simple isotropic expansion of the structures responsible.}
\label{fig:fig3}
\end{figure}

Figure \ref{fig:fig3} shows the photon energy at peak reflectance as a function of KH$_{2}$PO$_{4}$ molarity for the spectra shown in Figure \ref{fig:fig2}, which shows a remarkable linear trend. Figure \ref{fig:fig3} also shows the Gaussian half width as a function of soaking concentration for the $1\,$mm diameter spot over which reflectance was measured. In contrast to the trend noticed by the eye's color sensitivity, plotted in terms of photon energy the Gaussian half width increases somewhat with energy until broadly saturating above yellow.

The results shown in Figures \ref{fig:fig2} and \ref{fig:fig3} illustrate the possible use of biological photonic crystal structures as tunable, farm-grown, reflector arrays with minimal need for post-processing to achieve the desired peak reflectance wavelength. With such a simple method and a keen interest in industrial applications, one might ask, how durable are these structures? After all, they are biological structures and carcasses tend to decay rather quickly. In fact, we found that the color stripe showed a significant decay in about 15 minutes without any post-processing, where the color stripe turned a pale white without any color along it in about 2 hours. We suspect that biological decay processes cause the loss of color, where results may vary depending on the organisms and enzymes initially present that are involved in the decay.

\begin{figure*}[t]
\centering\includegraphics[scale=1]{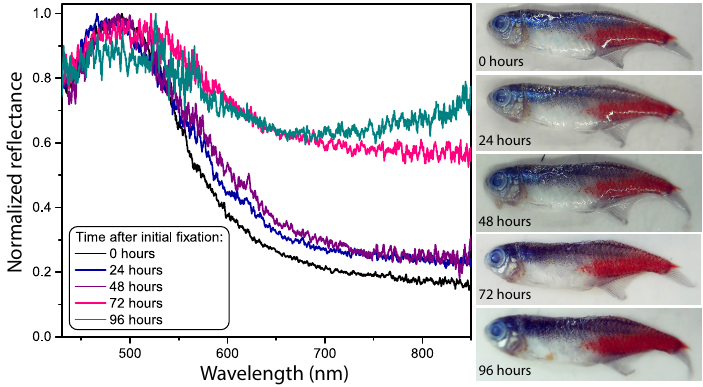}
\caption{The normalized reflectance as a function of time for half of a neon tetra fish in a 3\% formalin and $0.15\,$M KH$_{2}$PO$_{4}$ solution. The time starts directly after the final fixation step.}
\label{fig:fig4}
\end{figure*}

To maintain the brilliance of the neon tetra color stripe after death, formaldehyde was used in a post-process fixation method. Early attempts to use Karnovsky solution for TEM measurements led us to quickly remove all dehydrating chemicals during fixation. We found that even small concentrations of ethanol would rapidly turn the lateral color stripe on neon tetra carcasses into a pale white strip. We then developed a purely formalin-based fixation recipe to lengthen the degradation time of the lateral color stripe. We found that after an short initial formalin infiltration step at 5\% aqueous solution, lower formalin concentrations yielded optimum results for color stripe longevity. Figure \ref{fig:fig4} shows the reflectance of the color stripe measured in the ``neck'' region (immediately before the  \textit{operculum} as measured from the tail) for increasing time durations in $3$\% formalin, $0.15\,$M KH$_{2}$PO$_{4}$ solution after the initial fixation protocol which is detailed in the previous section.

\begin{figure*}[t]
\centering\includegraphics[scale=1]{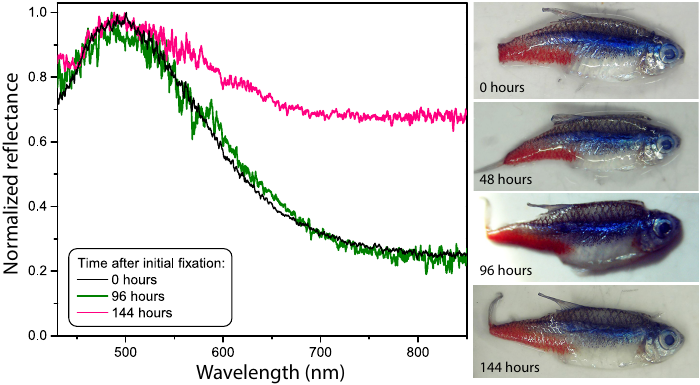}
\caption{The normalized reflectance as a function of time for half of a neon tetra fish stored in a formalin-free $0.15\,$M KH$_{2}$PO$_{4}$ solution. The time starts directly after the final fixation step.}
\label{fig:fig5}
\end{figure*}

We found that the color stripe's decay was reduced when introduced to an initial fixation formula followed by a 3\% formalin bath. The preferred method of fixation and preservation of biological tissue often requires the formalin concentration to remain the same during storage. Using low-concentrations of formalin for storage, we observed a lower decay rate of the structural color reflected from the lateral color stripe. The brilliance and color, however, only lasted a few days. A significantly longer lifetime should be required to use the material as Bragg reflectors in any industrial application.

Storage in a more standard concentration of formalin works well for maintaining many biological features; however, the reflective platelets in fish iridophores are guanine molecular crystals. Every guanine molecule in the crystal has a primary amine functional group. Aldehydes work to fix biological tissue by reacting with an amino acid's primary amine. Because guanine is an amino acid with a primary amine functional group, the platelets slowly degrade in the presence of molecules with the reactive aldehyde functional group. Because the platelets are crystalline and that free aldehydes will have a preference to react with amorphous structures during the fixation process, we hypothesized that a short fixation protocol followed by solution storage would reduce degradation of the color stripe while increasing the longevity of the biological tissue. Figure \ref{fig:fig5} shows the reflection spectrum of the iridophore Bragg reflector array when the tissue is fixed with formalin followed by storage in a KH$_{2}$PO$_{4}$ solution. The results provide a visual example of how formalin fixation followed by a solution absent in formalin can significantly increase the lifetime of amino acid crystal structures found in cells.

\begin{figure*}[t]
\centering\includegraphics[scale=1.3]{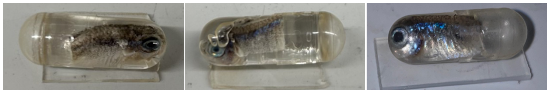}
\caption{Three neon tetra halves embedded with glycol methacrylate polymer. The three samples show differing amounts of iridescence in the color stripe region after polymerization. The sample on the left shows very little color remaining while the sample on the right retained most of its iridescence.}
\label{fig:fig6}
\end{figure*}

As shown in Figure \ref{fig:fig5}, the Bragg reflector arrays stored in KH$_{2}$PO$_{4}$ solution after fixation appear to be more stable. This increase in degradation lifetime could make this method a viable option for short-term applications that can conclude within a week. In this form, however, samples need to kept hydrated. Additionally, finding a long-lifetime form of Bragg reflector arrays sourced from fish could allow them to be used in a broader class of devices. Solid-state structures are commonly used in applications involving Bragg reflectors and associated devices. From this realization, we attempted to fabricated solid-state Bragg reflectors by embedding neon tetra iridophore networks in solid plastic. From previous attempts at fabricating thin slices of tissue for TEM analysis, we noticed that the preferred method of embedding tissue with Spurr resin after dehydration by ethanol resulted in structures that did not maintain any structural color. We later attempted to use a water-based, low-acid, glycol methacrylate from kits purchased from SPI Supplies. We were unable to complete the TEM analysis because of crumbling of the sample during the ultramicrotome steps. There was a notable feature, however, when using the water-soluble polymerization technique -- much of the structural color in the neon tetra color stripe was preserved. Figure \ref{fig:fig6} shows samples of neon tetra tissue embedded in glycol methacrylate. There were significant variations in the quality of the color stripe using the same preparation method on different neon tetra samples. We also tried to add KH$_{2}$PO$_{4}$ during the low-acid GMA infiltration steps. The addition of KH$_{2}$PO$_{4}$ did not inhibit polymerization, and the transparency of the glycol methacrylate polymer appeared to be unaffected; however, this addition to the procedure did not improve the variability in quality. No visual signs of decay were observed after polymerization. The samples, however, were placed in a dark cabinet and no high-intensity ``work horse'' studies have been performed on the samples. It appears that these samples are available for low-light intensity applications, where further study is necessary to determine their viability in high-intensity devices.

\section{Conclusion}

We conducted a preliminary evaluation of neon tetra iridophores in the lateral color stripe for potential \textit{ex vivo} applications. We showed broad, deterministic, and fast tunability of the peak optical reflectance could be easily achieved by exposure to KH$_{2}$PO$_{4}$ aqueous solutions. We also note that the width of the reflection band increased with an increase in the peak wavelength of reflection. The observed slow dispersion of the reflection band is consistent with changes in the heterogeneous spacings of cytoplasm between guanine platelets inside single iridophores and variations of peak reflections between neighboring platelets using the post-processing methods presented in this paper.

A protocol for formalin fixation was established to decrease the decay rate of iridophore reflection. We found that a relatively short formalin fixation protocol followed by storage in KH$_{2}$PO$_{4}$ aqueous solution yielded the best result for material lifetime. Further investigation into the plastization of tissue to form a solid-state Bragg reflector array showed that infiltration and embedding with glycol methacrylate resulted in stable samples with reflection bands being maintained in the ``neck'' region of the fish.

\section*{Funding}
This material is based upon work supported by the National Science Foundation under Grant No. 2337595.

\section*{Acknowledgments}
Nathan Dawson thanks Tina Carvalho at the University of Hawaii, Manoa, Biological Electron Microscope Facility for assistance and training to collect TEM images of neon tetra guanine crystal platelets. Nathan Dawson also thanks Prof. Derek Henderson for donating high-purity DI water, Prof. Matt Bohm for his help in using the Florida Polytechnic 3D printing facilities, and Prof. Ajeet Kaushik for use of the bio-safety cabinet in his laboratory.


\end{document}